\documentclass[%
 reprint,
 amsmath,amssymb,
 aps,prl
]{revtex4-1}

\usepackage{graphicx}% Include figure files
\usepackage{dcolumn}% Align table columns on decimal point
\usepackage{bm}% bold math
\usepackage{color}
\usepackage[colorinlistoftodos]{todonotes}
\begin{document}

\title{Orbital angular momentum transfer to stably trapped elastic particles in acoustical vortex beams}% Force line breaks with \\

\author{Diego Baresch}
\affiliation{
Sorbonne Universit\'e, CNRS UMR 7588, Institut des NanoSciences de Paris, INSP, F-75005, Paris, France.}%
 \altaffiliation[Currently at ]{Department of Chemical Engineering, Imperial College London.}
\email{d.baresch@imperial.ac.uk}

\author{R\'egis Marchiano}
\affiliation{Sorbonne Universit\'e, CNRS UMR 7190, Institut Jean le Rond d'Alembert,  F-75005, Paris, France.}

\author{Jean-Louis Thomas}%
\affiliation{
Sorbonne Universit\'e, CNRS UMR 7588, Institut des NanoSciences de Paris, INSP, F-75005, Paris, France.}%
\email{jean-louis.thomas@upmc.fr}


\begin{abstract}
The controlled rotation of solid particles trapped in a liquid by an ultrasonic vortex beam is observed. Single polystyrene beads, or clusters, can be trapped against gravity while simultaneously rotated. The induced rotation of a single particle is compared to a torque balance model accounting for the acoustic response of the particle. The measured torque ($\sim 10$~pNm for a driving acoustic power $\sim 40$~W/cm$^2$) suggests two dominating dissipation mechanisms of the acoustic orbital angular momentum responsible for the observed rotation. The first takes place in the bulk of the absorbing particle, whilst the second arises as dissipation in the viscous boundary layer in the surrounding fluid. Importantly, the dissipation processes affect both the dipolar and quadrupolar particle vibration modes suggesting that the restriction to the well-known Rayleigh scattering regime is invalid to model the total torque even for spheres much smaller than the sound wavelength. The findings show that a precise knowledge of the probe elastic absorption properties is crucial to perform rheological measurements with manoeuvrable trapped spheres in viscous liquids. Further results suggest that the external rotational steady flow must be included in the balance and can play an important role in other liquids.
\end{abstract}

\maketitle
Since the demonstration of particle trapping and manipulation of transparent particles by a single focused laser beam by Ashkin \textit{et al}\cite{Ashkin86}, ``optical tweezers" that can pull a trapped particle in all three directions have found numerous applications, particularly in the biophysical research \cite{Ashkin87Nature,Svoboda1994}. Using the radiation pressure of sound, rather than light, it was recently demonstrated that "acoustical tweezers" could operate as three dimensional traps for elastic particles using a single ultrasonic vortex beam first numerically \cite{BareschJAP} and then experimentally \cite{BareschThese,Baresch2016}. The change in nature of the propagating wave presents several advantages for contactless manipulation as the possibility to operate through turbid media, allow penetration in tissue, largely increase the magnitude of the trapping force and the size of the particles. 

The attraction in the direction of the intensity gradient of transparent, dielectric, objects in optical tweezers relies on the transfer of the momentum carried by photons. It is however well established that photons can also carry angular momentum and exert torques \cite{Beth1936}. This important degree of freedom has proven important for the controlled rotation of optically trapped particles with spin or orbital angular momentum (OAM) of photons \cite{Marston1984,He1995,Simpson1997,Friese1998,Bishop2004}. 

In contrast, longitudinal acoustic waves in liquids do not carry momentum \cite{McIntyre1981}, but instead can induce a mean stress, after exchange of a flux of momentum -- \textit{e.g.} by scattering or absorption -- either it be linear or angular. Rayleigh first analyzed and quantified the torque exerted on a disk suspended in a sound field\cite{Rayleigh1882}. In that particular case the torque is understood as a consequence of the uneven radiation pressure exerted on the surface of a disk unaligned with the sound propagation direction \cite{Maidanik1958} in a way that any object of irregular form could experience a net radiation torque. 
Accounting for the finite size of the viscous boundary layer $\delta=\sqrt{\frac{2\mu}{\rho\omega}}$ around an object's surface, where $\omega$ is the pulsation, $\mu$ the dynamic viscosity and $\rho$ the density of the suspending fluid respectively, the elliptical motion of fluid particles in a system of out-of phase orthogonal standing waves can induce the rotation of axisymmetric objects  \cite{Wang1977,Busse1981}. Combined with acoustical levitation \cite{Apfel81,Trinh85}, systems of counter-propagating waves have been selected as an advantageous method to induce the rotation of matter in air \cite{Biswas1991,Foresti2014} and are at the basis of the rotation of spherical, cylindrical and anisotropic particles, including cells, in fluids \cite{Shwarz2013,Lamprecht2015,Hahn2016,Bernard2017}.

The momentum flux vector, or Poynting vector, of an acoustical vortex (AV) will locally point in the direction of the helicoidal wavefront offering an additional degree of freedom under which it will be exchanged: the OAM of sound. The direct OAM transfer to matter has been observed in air \cite{Volke-Sepulveda2008,PadgettAcous} and water \cite{Demore2012,Anhauser2012,Hong2015} through absorption or chiral scattering \cite{Wunenburger2015}. AVs have recently been used to simultaneously levitate and rotate particles in air \cite{Marzo2015} but the lack of viscosity leads, however, to an off-axis rotational instability that can be controlled at the expense of the decrease of the net OAM transfer \cite{Marzo2018}. Nonetheless, the physical mechanisms driving the acoustic torque is unclear. No absorption processes were considered, suggesting that the main mechanisms leading the particle to spin around its axis were overlooked.   Additionally, the demonstration of the co-existence of the axial negative gradient force \cite{Baresch2016} and driving torque is not evident. A negative gradient force, pulling a particle against the acoustical momentum flux, is a crucial feature in the development of acoustical tweezers. In liquids, acoustic radiation forces and torques have experimentally been shown to exceed by 6 orders of magnitude their optical counterpart \cite{Anhauser2012,Demore2012,Thomas2017}. Hence, despite the significant potential of combining simultaneous trapping and rotation with a single-beam for selective particle manipulation, a quantitative experimental test of acoustical OAM transfer models is still missing.

\begin{figure}
\centering \includegraphics[width=0.8\linewidth]{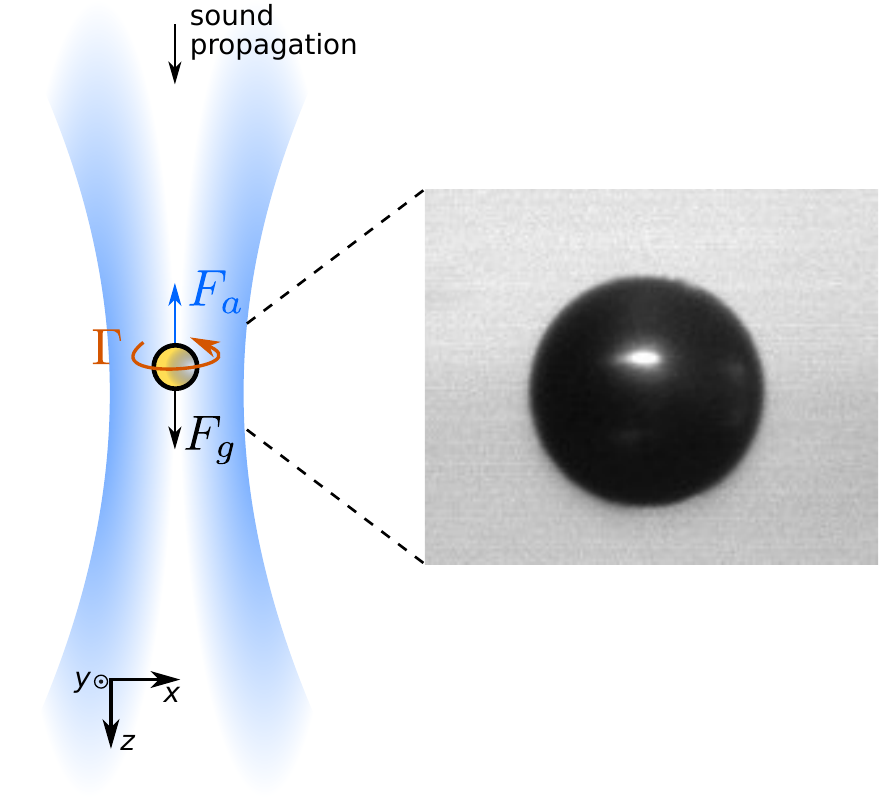}
\caption{(Color online) Schematic view of the ultrasonic vortex beam trapping a janus polystyrene particle coated with thin layer of gold. The negative acoustic gradient force, $F_a$, balances the gravity force $F_g$. The transfer of OAM from the beam to the particle and fluid bulk applies a torque $\Gamma$ driving an on-axis rotation. A photograph of a trapped and rotating particle is also shown.}
\label{fig1}
\end{figure}

In this Letter, we report on the observation of the simultaneous rotation and trapping of particles pulled by a negative gradient force. This is done by using AV-based acoustical tweezers to stably trap janus polystyrene particles (see Fig.\ref{fig1}) and measuring their rotation rate. The steady spinning frequency is used to derive a torque balance based on analytical calculations of the non linear acoustical torque. We find that, for a single spherical polystyrene bead, it is crucial to fully model the elastic scattering beyond the usual long-wavelength regime that is found to be invalid for the acoustic torque, within the usual size bonds set for this limit. For a particle size of $a/\lambda \sim 0.13$ where $\lambda$ is the driving wavelength, the dissipation in the solid particle bulk and viscous boundary layers involved in the dipolar and quadrupolar oscillations are responsible for the OAM transfer and driving torque. Further observations demonstrate the stable trapping and controlled rotation of asymmetric clusters of particles.  

\paragraph{Experimental setup}
Figure \ref{fig1} shows a schematic of the setup and a photograph of a trapped and spinning particle. Using the experimental setup previously described \cite{Baresch2016}, a focused AV (frequency $f=1.15$MHz, wavelength $\lambda=c_0/f = 1.3$~mm for a speed of sound $c_0=1480$~m/s at 20 degrees Celsius) exerts a lateral force trapping the particle in its zero pressure core and the tight focusing gives rise to the negative pulling force operating against gravity (Movie 1 in SI). Polystyrene beads (Polysciences Inc., of radius $2a\sim$360$\mu$m) were partially coated with a nanometric layer of gold to optically observe their rotational motion at a rate of 159 frames per second. The acoustic properties of the particles are unaltered by the coating procedure. The OAM of an AV with topological charge $m$ (phase variation $e^{im\varphi}$ around the $z$ axis) is related to the total energy density $E$ by $M=\frac{m}{\omega}\langle E \rangle$, \cite{Thomas2003}. It is thus possible to increase the torque by increasing the value of $m$. An immediate consequence is, however, the reduction of the linear momentum. We have previously analyzed theoretically that this translates in the reduction of the stiffness of the acoustic trap in both axial and lateral directions \cite{BareschJAP}. Hence, for the purpose of demonstrating simultaneously pulling against gravity and spinning in a context where gravity plays an important role, vortex beams with the minimal topological charge $m=\pm 1$ are generated. The maximum acoustic pressure used is $p_0=0.8\pm0.1$ MPa measured on the vortex ring or equivalently an intensity of 42 W/cm$^2$. Note that higher order AVs with $m=3$ and $5$ were recently used in air to levitate expanded polystyrene particles \cite{Marzo2018}. However, only the acoustic pushing (positive) force was observed to counteract the pull of gravity, the acoustic negative gradient force being insufficient as predicted theoretically \cite{BareschJAP}. Thus, the current failure of using high order AVs in single-beam tweezers is so far common to both air and water.

\begin{figure}
\centering \includegraphics[width=1.0\linewidth]{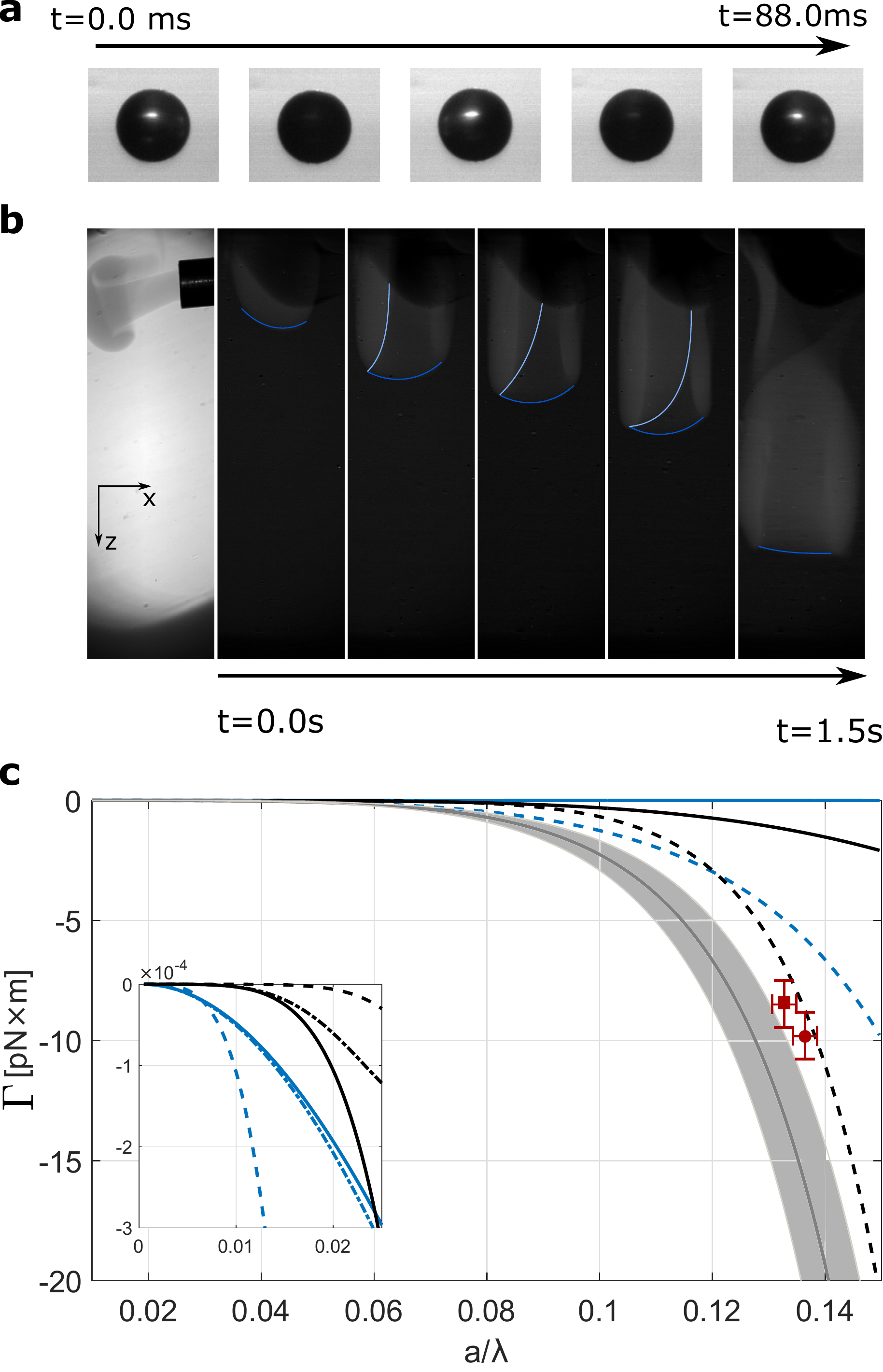}
\caption{OAM transfer to a stably trapped particle. In a) snap shots are shown of one complete revolution of a polystyrene particle with $a=174\mu$m. b) Rotational acoustic flow generated by the absoption of the AV in the fluid bulk. An injected drop of ink allows for the observation of the developping flow upon emission of the AV. c) Computation of the acoustic torque as a function of the particle radius and particle vibration modes (see text). The measured torque with error bars (see text) are added for two independent rotation experiments with spherical probes having $a=172$ and $174\mu$m respectively.}
\label{fig2}
\end{figure}

\paragraph{OAM transfer balance}
Upon starting the AV emission, two different polystyrene particles of respectively $a=172 \pm 4 \mu$m  and $174 \pm 4 \mu$m radius are accelerated towards their equilibrium position where all forces balance out. The OAM transfer of a vortex beam of topological charge $m=-1$ results in the rotation at a rate of $f_r=10.5$ and 11 Hz respectively (Fig.\ref{fig2}a). The spinning rate is detected by the optical mean gray value extracted from each frame of a video and displayed (Methods in SI). The dark side of the particle appears twice in a revolution when the golden face lies in between the light source and the camera (see Movie 1 in SI). The angular speed  $\Omega=2\pi\times f_r\simeq 70$ rad/s suggests that the driving acoustic torque $\Gamma$ on the spheres is balanced by the drag  $\Gamma_D=-8\pi \mu a^3 \Omega\simeq -10$ pN.m acting in the opposite direction in a fluid of density $\rho=1000$~kg/$m^3$ and viscosity $\mu=1 mPa\cdot s$ for which the low Reynolds number approximation Re$=a^2\Omega/\nu\ll 1$ holds. Note that inertial effects acting in a time scale of a few milliseconds \cite{Lamprecht2015} are unresolved with this setup. 

Among the possible means to induce an OAM transfer, Anh\"auser \textit{et al} invoked the rotational flow induced in the bulk of a viscous mixture of aqueous glycerol by the absorption of an incident AV \cite{Anhauser2012}. Though water has a much lower viscosity, here we were able to directly observe the flow by injecting a solution of ink for which the diffusion time scale was much shorter than for the acoustically forced flow. In fig \ref{fig2}b, we can follow the temporal evolution of the flow in the focal zone by looking at the evolution of the ink concentration (Movie 2 in SI). It reveals the simultaneous axial and rotational components of the flow field that exists without the presence of the sphere. The rotational flow can be evaluated (See Supplemental Material) to be approximately $u_\varphi \sim 1~$mm/s in the vicinity of the trapped sphere, weaker than the axial flow, $u_z \sim 4~$mm/s. While devising a method to measure the fine features of this complex 3D flow was beyond the scope of this study, we infer that it contributes to weakly reduce the drag on the sphere, introducing an approximate error on the torque $\Gamma_D'= 8\pi \mu a^3(u_\varphi/a)\simeq 1$ pN.m.    

The mechanical torque driving the rotation is the time-averaged acoustic torque exerted on the particle partly scattering and absorbing the incident OAM. As described in Supplemental Material, we compute the torque $\Gamma$ exerted by the focused AV on a polystyrene bead in water with density $\rho_p=1050~$kg/m$^3$ and speed of sound  $c_\ell^\infty=2350~$m/s and  $c_t^\infty = 1100~$m/s for longitudinal and transverse internal waves respectively. Polystyrene, as other amorphous solids with a glass transition, has a visco-elastic behavior well modeled as a Maxwell material. Thus the bulk and shear absorption coefficients are frequency dependent and ultrasonic measurements have been reported in the literature \cite{Takagi2007}. The longitudinal absorption coefficient $\alpha_\ell =30~$Np/m is found to be roughly three times weaker than its transverse counterpart $\alpha_t = 100~$Np/m. We also account for losses involved in the boundary layer supporting shear viscosity waves \cite{Allegra,Settnes2012}. This dissipation will also give rise to a steady mean flow, originally analysed by Schlichting \cite{Wiklund2012}. It's influence in calculating the torque is discussed in Supplemantary Material. The field scattered by the particle can be expanded in terms of multipoles, of which we find a significant influence of the dipole and quadrupole. The monopolar mode is annihilated as a consequence of the broken symmetry by the incident vortex field \cite{Baresch2016}. In figure \ref{fig2}c, we plot the different contributions to the torque as a function of $a/\lambda$. The torque arising from the dissipation of the dipole is shown with solid curves and with dashed curves for the quadrupole when only the viscous dissipation in the fluid (blue) or absorption in the particle (black) are considered independently. The total torque is in turn the sum of all four curves. Allowing for a 14$\%$ hydrophone uncertainty on the AV's maximum pressure amplitude ($p_0=0.8\pm0.1$ MPa), we set bounds for the theoretical evaluation of the total torque with the arising error (gray area). The two torque measurements are also shown (red square and circle for $a=172$ and $174\mu$m respectively) with a corresponding $\pm 4\mu$m (2 pixels) error in evaluating $a$ and $\pm 1$~pNm in evaluating the torque motivated by the observed drag reduction discussed previously. Spanning the range of investigated sizes in experiment would require additional spherical probes with identical absorption that are difficult to obtain commercially. Furthermore, smaller probes will additionally be pushed down the trap by the axial Stokes drag ($\propto a$) until they escape when the negative pulling force fails ($\propto a^3$). The different axial positions of smaller probes would require to recast the estimation of the torque for the divergence of the incident focused beam.

\begin{figure*}[ht]
\centering \includegraphics[width=1.0\linewidth]{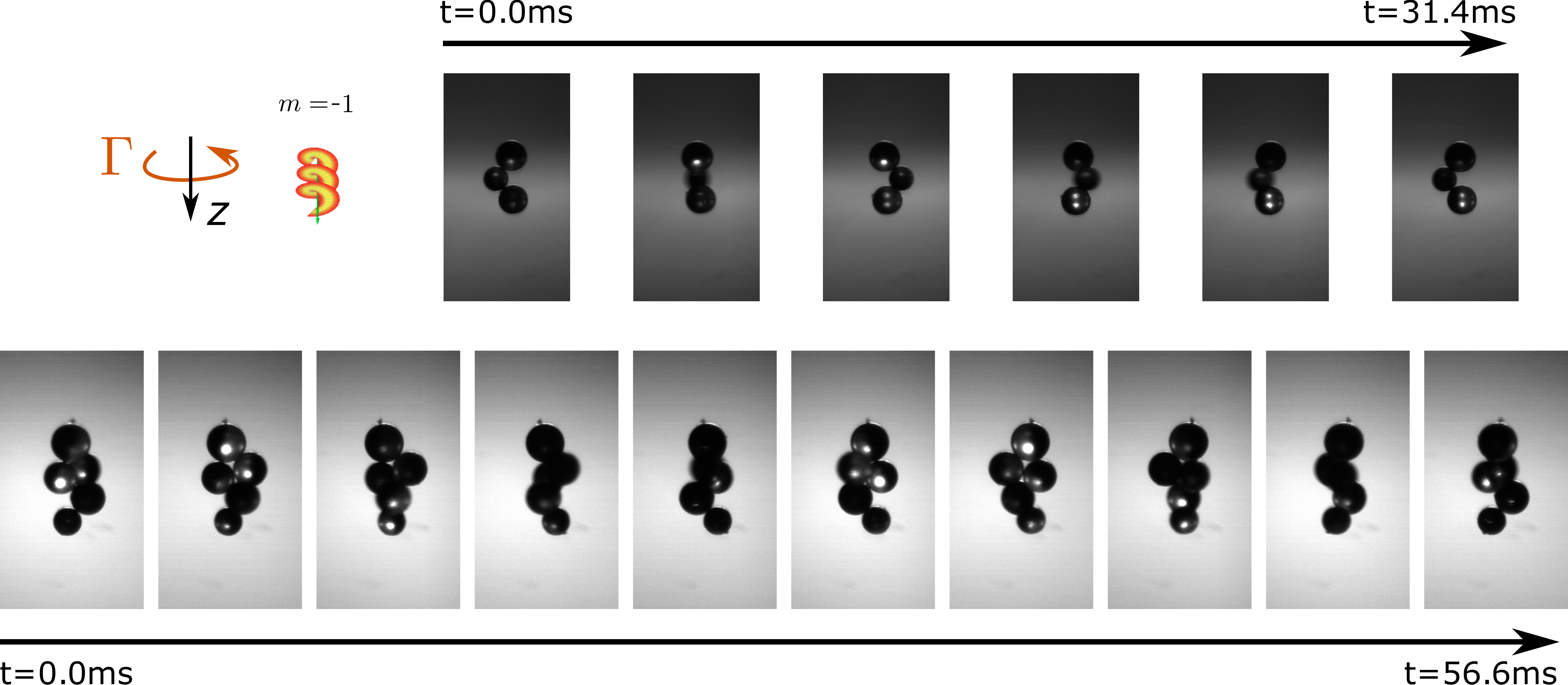}
\caption{(Color online) Simultaneous trapping and rotation of clusters formed by three or five polystyrene particles. The particle sizes are 352, 280 and 320 $\mu$m (top to bottom) and 380, 320, 320, 320 and 268 $\mu$m (top to bottom).}
\label{fig3}
\end{figure*}

Our measurements are in good quantitative agreement with our calculations that do not include any adjustable parameter. In the best case (error compensations) the measurement and calculation are in agreement within less than $5\%$. However the error can reach near 100\%  (no compensation) as a consequence of the systematic 14$\%$ error in measuring $p_0$ that affects quadratically the torque and in light of the high contribution of the sphere's visco-elastic absorption (black curves) that rapidly dominates the torque balance and our uncertainty in ascribing the value for the absorption coefficients. Indeed, we did not find additional data in the literature for $\alpha_\ell$ and $\alpha_t$. Therefore, measuring their value for our spheres could largely improve the torque balance. Importantly, the fact that absorption within the sphere is the dominating driving mechanism contrasts with previous results restricting the description to the viscous boundary layer around the sphere \cite{Lamprecht2015,Bernard2017} or neglecting absorption processes entirely \cite{Marzo2018}. Finally, within the particle size range considered theoretically, the large contribution of the quadrupolar oscillation mode, even for $a/\lambda<0.05$ (see the close view in Fig.\ref{fig2}c), is in contradiction with the common assumption that it can be neglected in the long wavelength (Rayleigh) scattering regime ($a/ \lambda\ll 1$). Though recent analytical calculations under these conditions are here recovered for boundary layer dissipation \cite{Zhang2014} and particle visco-elastic absorption \cite{Silva2014} of the dipolar oscillation (dash-dotted curves in the close view), the dissipation of the quadrupolar mode leads to a torque that rapidly dominates as the radius of the particle increases and must be considered in the full torque evaluation even for small $a/\lambda$ ratios. For example, taking a bond that could generally be considered as lying in the long-wavelength regime, $a/\lambda=0.02$, our torque estimation is nearly 10 times larger than the one predicted in that limit when boundary layer dissipation and particle absorption are considered separately (Eq.13 in \cite{Zhang2014} and Eq.30 in \cite{Silva2014}), mainly as a consequence of the viscous dissipation of the quadrupole vibrating in a viscous fluid (See the Supplemental Material for an extended discussion).

\paragraph{Particle clusters and rotation control}
We additionally observed the rotation of clusters of three or five particles (Fig.\ref{fig3}). The rotation rate of the first is measured to be 32Hz and 18Hz for the latter. This superior rotation rate compared to the single particle case suggests that the clusters' asymmetry involves an additional mean to transfer OAM. The asymmetry in the scattered field leads to a torque that has been observed in 2D cases for glass fibers \cite{Shwarz2013} or biological cells \cite{Bernard2017}. We note that the existence of a stable trapping potential for clusters of these dimensions is unexpected from single particle theories of acoustic radiation forces. Accordingly, theses observations call for a deeper understanding of the interaction of AVs with various particles involving multiple scattering processes and secondary forces and torques \cite{BruusMS,Silva2016}.   

Finally, as reported in a preliminary form in \cite{BareschThese} in water and recently for AVs in air \cite{Marzo2018}, it is possible to fine tune the rotation rate of the trapped particles by rapidly switching the handedness of the incident AV (Movie 3 in Supplementary Materials). The strategy benefits from the two different time scales involved in the wave phenomena. The fast time scale is determined by the acoustic wave oscillations defined by the acoustic period $T=1/f=1\mu$s. The mean torque arising from the OAM transfer builds up over this time scale. The slow time scale, $\tau$, will ultimately depend on the viscosity of the host fluid and the strength of the driving torque leading to the steady rotation rate. We find that by alternating the wavefront handedness at a time scale $T_0 =400\mu$s (such that $T \ll T_0 \ll \tau\sim 1$~ms), it is possible to fine tune the rotation rate from the maximum rate available - determined by that obtained at a fixed power with a single handedness - to nearly zero when during the time duration $T_0$, the topological charge is set to $m=1$ half of the time and to $m=-1$ the other half. When the charge is set to be $m=1$ during a period equal to $0.75 T_0$, the rotation speed is half of what is obtained with a single handedness during the whole period $T_0$. Remarkably there is no alteration of the trapping position given that the gradient force is invariant with the sign of $m$.

AV based acoustical tweezers can transfer a controllable amount of OAM to 3D trapped elastic particles. The rotation is driven primarily by the viscous dissipation within a thin viscous boundary layer in the liquid around the particle and the absorption in the particle itself. Both the dipolar and quadropolar vibration modes are dissipated suggesting that the full calculation of the scattered field must include the visco-elasticity of the trapped material and the viscous boundary layer $\delta=(2 \mu /\rho \omega)^{1/2}$. Our results suggest that: (1) the existence of a theoretical long-wavelength regime neglecting the quadrupole can be difficult to translate in experiment, (2) the $(a/\lambda)$ bond for this regime can be material dependant and finally (3), both absorption in and around the sphere are crucial contributions. While the host medium was here restricted to water, the important role of the rotational flow induced by the viscous attenuation of the AV in the liquid bulk should be considered in evaluating the force and torque in other viscous fluids or at higher frequencies \cite{Eckart1948,Riaud2014,Riaud2017}. We believe that stable 3D trapping combined with tunable rotation rates has potential for applications as contactless micro-assembly and \textit{in situ} rheology of small fluid volumes and elastic particles in viscous and complex fluids.

\bibliography{Force}

\end{document}